\documentclass[prd,preprint,superscriptaddress,showpacs,nofootinbib,%
tightenlines ]{revtex4}
\usepackage{epsfig}
\usepackage{amssymb}

\newcommand{\ben}{\begin{displaymath}}
\newcommand{\een}{\end{displaymath}}
\newcommand{\be}{\begin{equation}}
\newcommand{\ee}{\end{equation}}
\newcommand{\bea}{\begin{eqnarray}}
\newcommand{\eea}{\end{eqnarray}}
\newcommand{\nn}{\nonumber \\ }

\begin{document}
\preprint{FZJ-IKP-TH-2009-16}
\preprint{HISKP-TH-09/18}
\preprint{MKPH-T-09-10}

\title{Regularization, renormalization and ``peratization'' in effective field
  theory for two nucleons}
\author{E.~Epelbaum}
\affiliation{Forschungszentrum J\"ulich, Institut f\"ur Kernphysik (Theorie) and
  \\J\"ulich
Center for Hadron Physics, \\D-52425 J\"ulich, Germany}
\affiliation{Helmholtz-Institut f\"ur Strahlen- und Kernphysik (Theorie) and
  \\ Bethe
Center for Theoretical Physics, Universit\"at Bonn, \\ D-53115 Bonn, Germany}
\author{J.~Gegelia}
\affiliation{Institut f\"ur Kernphysik, Johannes
Gutenberg-Universit\"at, \\ D-55099 Mainz, Germany}
\affiliation{High Energy Physics Institute,
Tbilisi State University, \\0186 Tbilisi, Georgia}
\begin{abstract}
We discuss conceptual aspects of renormalization in the context of effective
field theories for the two-nucleon system. It is shown that, contrary to
widespread belief, renormalization scheme dependence of the scattering
amplitude can only be eliminated up to the order the calculations
are performed. We further consider an effective theory for an exactly solvable
quantum mechanical model which possesses a long- and short-range
interaction to simulate
pionful effective field theory. We discuss the meaning of low-energy
theorems in this model and demonstrate their validity in calculations with a
finite cutoff $\Lambda$ as long as it is chosen of the order of the
hard scale in the problem. Removing the cutoff by taking the limit
$\Lambda \to \infty$ yields a finite result for the scattering amplitude
but violates the low-energy theorems and is, therefore, not compatible
with the effective field theory framework.

\end{abstract}
\pacs{11.10.Gh,12.39.Fe,13.75.Cs}

\maketitle

\section{Introduction}
\label{introduction}
Almost two decades ago Weinberg proposed a way to extend baryon chiral
perturbation theory to few-nucleon systems \cite{Weinberg:rz,Weinberg:um}.
These seminal papers triggered an intensive research activity
starting with the pioneering work of Ref.~\cite{Ordonez:1992xp}.
In this approach, chiral perturbation theory is applied to the effective
potential, defined as the sum of all possible $N$-nucleon irreducible
diagrams, rather than to the scattering amplitude. The amplitude is then generated
by solving the corresponding dynamical equation such as the Lippmann-Schwinger
(LS) equation in the two-nucleon sector. For recent reviews and references
the reader is referred to
Refs.~\cite{Bedaque:2002mn,Epelbaum:2005pn,Epelbaum:2008ga}.

While phenomenologically successful, the consistency of Weinberg's approach
was questioned by several authors.  The nucleon-nucleon (NN)
potential in this formalism is non-renormalizable in the traditional sense,
i.e.~iterations of the LS
equation generate divergent terms with structures which are not included in the
original potential. For example, the leading-order (LO) NN potential is given
by derivative-less contact interactions contributing only to S-waves and
the one-pion exchange (OPE) term whose spin-triplet part behaves at short
distances as $1/r^3$ and, therefore, generates divergences also in higher partial
waves. Consequently,
renormalization of the solution
of the LS equation requires inclusion of contributions of infinitely many
higher-order short-range operators in the potential (counterterms).
The freedom in the choice of the finite parts of counterterms is compensated
by the running of the corresponding renormalized coupling constants.
It has been argued \cite{Kaplan:1996xu} that the coefficients in front of the
divergent parts of the
counterterms contributing at a given order set the scale of the corresponding
renormalized couplings. As a consequence, even if these couplings were natural
at some value of the renormalization scale, they would become unnaturally large
for slightly different values of this parameter.
This problem, also treated non-perturbatively, is what is usually referred to
as inconsistency of Weinberg's approach, see also Ref.~\cite{Beane:2001bc} for
a related discussion.

An
alternative power counting scheme has been proposed by Kaplan, Savage and Wise
(KSW) \cite{Kaplan:1996xu,Kaplan:1998tg,Kaplan:1998we,Savage:1998vh},
in which the troublesome OPE contribution to the potential is
shifted from LO to next-to-leading order (NLO). The
LO dynamical equation becomes renormalizable, both perturbatively and non-perturbatively,
i.e. all divergences
can be absorbed into redefinition of low-energy constants (LECs)
entering the potential. Moreover,
the LO equation is exactly solvable and dimensional regularization
can be applied. All corrections are treated perturbatively which guarantees
that {\it all} divergences are absorbed into redefinition of
parameters entering at a given order. 
Unfortunately the resulting perturbative
expansion for the scattering amplitude was found not to converge for nucleon
momenta of the order of the pion mass at least in certain spin-triplet channels
\cite{Fleming:1999ee}, see however, Ref.~\cite{Beane:2008bt} for a new formulation
which is claimed to yield a convergent expansion. The reason for the breakdown
of the KSW expansion was attributed to the perturbative treatment of the
pion-exchange contributions
\cite{Gegelia:1998ee,Gegelia:1999ja,Cohen:1998jr,Cohen:1999iaa,Cohen:1999ds}.
This appears to be in line with phenomenological successes of Weinberg's approach
which treats pion exchange contributions  nonperturbatively. Indeed, the most
advanced analyses of the NN system at next-to-next-to-next-to-leading order
in the Weinberg's power counting scheme demonstrate the ability to accurately
describe NN
scattering data up to center-of-mass momenta  at least of the order $\sim 2
M_\pi$ \cite{Entem:2003ft,Epelbaum:2004fk}. It is important to emphasize that
these studies are carried out within the cutoff EFT along the lines of
Lepage \cite{Lepage:1997,Lepage:1999kt,Lepage:2000} who argued that the
cutoff parameter in such calculations should be taken of the
order of the relevant hard large scale such as e.g. the mass of the $\rho$
meson, see also Refs.~\cite{Gegelia:gn,Gegelia:1998iu,Park:1998cu,
Epelbaum:2004fk,Gegelia:2004pz,Epelbaum:2006pt}.

The fairly narrow range of cutoffs $\Lambda = 450
\ldots 600$ MeV used in Refs.~\cite{Entem:2003ft,Epelbaum:2004fk} was
criticized by Nogga et al.~\cite{Nogga:2005hy} who considered low NN partial
waves based on the OPE potential and contact interactions employing a much
larger range of cutoffs with $\Lambda < 4$~GeV. They found that higher-order
counterterms have to be promoted to LO in the $^3P_0$, $^3P_2$-$^3F_2$ and,
possibly, the $^3D_2$ channel in order to stabilize the amplitude in the
employed cutoff range. The authors of Ref.~\cite{Nogga:2005hy} conjecture
that the ``mixture of perturbative treatment of higher partial
waves, resummation of lower partial waves, and promotion of a finite
number of counterterms is the most consistent approach'' to
chiral effective field theory (EFT) in the two-nucleon sector, see however, Ref.~\cite{Epelbaum:2006pt}
for criticism. The possibility of a perturbative treatment of two- and
more-pion exchange corrections to the potential was explored using
renormalization-group methods \cite{Birse:2003nz,Barford:2002je,Birse:2005um}.
Finally, the consequences of completely removing the cutoff
$\Lambda$ by
taking the limit $\Lambda \to \infty$ in the LS equation based on the NN
potentials at various orders in chiral EFT are also being explored by
several groups
\cite{Frederico:1999ps,PavonValderrama:2003np,PavonValderrama:2004nb,
Timoteo:2005ia,PavonValderrama:2005gu,
PavonValderrama:2005wv,PavonValderrama:2005uj,Higa:2007gz,Entem:2007jg,
Long:2007vp,Yang:2007hb,Valderrama:2008kj,Yang:2009kx}.

The purpose of this paper is to clarify some conceptual issues related to
renormalization in the context of EFT for the two-nucleon system. First,
we discuss renormalization scheme dependence of the scattering
amplitude in the KSW and Weinberg's approaches. Contrary to widespread
belief, we show that renormalization scheme independence in the KSW
framework is only achievable up to the order to which the calculations are
performed. From this point of view, the KSW framework does not offer any
conceptual advantage over the Weinberg's approach. Secondly, we regard cutoff
EFT and explore the consequences of completely removing (or taking very large
values of) the cutoff. To that aim, we construct effective theory for an
exactly solvable quantum mechanical model with long- ($r_l \sim
m_l^{-1}$) and short-range ($r_s \sim m_s^{-1} \ll m_l^{-1}$) interactions of
a separable type valid for momenta of the order $k \sim m_l$. This can be
viewed as a toy-model for pionful EFT. We explain the meaning
of low-energy theorems in this model using the KSW-like framework with
subtractive renormalization and
demonstrate their validity in the Weinberg-like approach with a finite cutoff
$\Lambda$ as long as it is chosen of the order $\Lambda \sim m_s$. Next, it is
shown that taking the limit $\Lambda \to \infty$ yields a finite result for
the amplitude but
leads to breakdown of low-energy theorems. This procedure is, therefore, not
compatible with the EFT framework.
We argue that $\Lambda$ should not be
taken (considerably) larger than the short-range scale $m_s$ in the problem.

Our paper is organized as follows. In section \ref{pionless} we consider
renormalization scheme dependence in the KSW and Weinberg's approaches
concentrating mainly on a pionless theory.
Cutoff EFT for the exactly solvable toy model is discussed in section
\ref{cutoffEFT}. Finally, the findings of our work are briefly summarized in
section \ref{summary}.

\section{KSW versus Weinberg's approach}
\def\theequation{\arabic{section}.\arabic{equation}}
\setcounter{equation}{0}
\label{pionless}

For very low energies the effective non-relativistic
Lagrangian relevant for S-wave nucleon-nucleon scattering can be
written as \cite{Weinberg:rz,Weinberg:um,Kaplan:1996xu}:
\be
{\cal L}  =  N^{\dagger}\bigg[ i\partial_t + {\nabla^2 \over
2m} \bigg] N - {C_S \over 2}\left( N^{\dagger}N\right)^2-{C_T\over 2}
\left( N^{\dagger}\mbox{\boldmath $\sigma$}N\right)^2 -{C_2\over 2}\left(
  N^{\dagger}\nabla^2 N\right) \left(
N^{\dagger}N\right)
+ \mbox{\,h.c.}+ \ldots \,,\label{e1}
\ee
where the nucleonic field $N$ is a two-component spinor in spin
and isotopic spin spaces and $\mbox{\boldmath $\sigma$} $
are the Pauli matrices acting on spin indices. Further, $m$ is the nucleon mass
and $C_T$, $C_S$ and $C_2$ are low energy coupling
constants. The LO contribution to the NN potential in the
${ }^1S_0$ partial wave is
\begin{equation}
V_0\left( {p},{p'}\right)=C_S-3\,C_T =C\,,
\end{equation}
while the NLO one has the form:
\begin{equation}
V_2\left( {p},{p'}\right)=C_2\left({p}^2+{p'}^2\right).
\end{equation}
In Weinberg's approach, the scattering amplitude is obtained by
solving the Lippmann-Schwinger equation. For the potential $V_0+V_2$, the
well-known solution for the on-the-energy-shell $T$-matrix reads, see e.~g.~\cite{Phillips:1997xu},
\be
T  =  \frac{C+C_2^2I^\Lambda_5+ k^2C_2\left( 2-C_2I^\Lambda_3\right)}{\left(
1-C_2I^\Lambda_3\right)^2-\left[C+C_2^2I^\Lambda_5+ k^2C_2\left(
2-C_2I^\Lambda_3\right)\right]\,I^\Lambda(k)}\,, \label{2}
\ee
with the cutoff-regularized loop integrals defined as
\begin{eqnarray}
I_n^\Lambda & = & -{m \over (2\pi )^3}\int {d^3l}\, l^{n-3}\,\theta \left(\Lambda-l
\right) = -\frac{m\,\Lambda^n}{2n\pi^2} \,,\; \mbox{
  with } \; n=1,3,5\,, \nn
I^\Lambda(k) & = & {m\over (2\pi )^3}\int {d^3l \,\theta \left(\Lambda-l
  \right)\over
k^2-l^2+i\eta} = I_1^\Lambda -\frac{i\,m\, k}{4\pi} -
\frac{m k}{4\pi^2} \,\ln \frac{\Lambda-k}{\Lambda+k}
 \,, \label{3}
\end{eqnarray}
where $k$ refers to the on-shell momentum in the NN center-of-mass system and
the last equation is valid for $k < \Lambda$.
To renormalize (\ref{2}) we divide loop integrals into the divergent and
finite parts and take the limit $\Lambda \to \infty$:
\begin{eqnarray}
I_n & \equiv & \lim_{\Lambda \to \infty} I_n^\Lambda = \lim_{\Lambda \to
  \infty} \left(I_n^\Lambda + \frac{m \mu_n^n}{2 n \pi^2}\right)
-\frac{m \mu_n^n}{2 n \pi^2} \equiv \Delta_n(\mu_n)+I_n^R(\mu_n)\,, \mbox{
  with } n=3,5\,,\nonumber\\
I(k) & \equiv & \lim_{\Lambda \to \infty}  I^\Lambda (k) = \lim_{\Lambda \to \infty}
\left(I_1^\Lambda + \frac{m \mu}{2 \pi^2}\right)
+\left[-\frac{m \mu}{2 \pi^2}  -\frac{i\,m\, k}{4\pi}  \right] \equiv
\Delta(\mu)+I^R(\mu,k)\,. \label{splitting2}
\end{eqnarray}
Here, $\Delta_n(\mu_n)$ and $\Delta(\mu)$ denote the divergent parts
of the loop integrals while $I_n^R(\mu_n)$ and $I^R(\mu,k)$ are finite. The
splitting of
loop integrals in Eq.~(\ref{splitting2}) is not unique. The freedom in the
choice of renormalization
conditions is parameterized by $\mu$ and $\mu_n$. The divergent parts
$\Delta_n(\mu_n)$
and $\Delta(\mu)$ are to be canceled by contributions of
counterterms. To absorb all appearances of  $\Delta_n(\mu_n)$
and $\Delta(\mu)$ in Eq.~(\ref{2}) one needs to include
contributions of an infinite number of counterterms of increasingly higher
orders in powers of momenta \cite{Gegelia:gn}.
While it is impossible to write down these counterterms
explicitly, one can take their contributions into account by
dropping $\Delta_n(\mu_n)$ and $\Delta(\mu)$ terms and replacing
$C$ and $C_2$ by renormalized couplings. The subtracted
(renormalized) amplitude reads:
\begin{equation}
T = \frac{C_R+(C_2^R)^2 I_5^R(\mu_5)+ k^2C_2^R\left(
2-C_2^RI_3^R(\mu_3)\right)}{\left( 1- C_2^R
I_3^R(\mu_3)\right)^2-\left[ C_R+(C_2^R)^2 I_5^R(\mu_5)+
k^2C_2^R\left( 2-C_2^RI_3^R(\mu_3)\right) \right]\,I^R(\mu,k)}\,.
\label{rentmp2}
\end{equation}
Note that we are free to fix the freedom in finite parts of loop
integrals. Any scheme that puts an effective cut-off of the order of
external momenta is equally good from the EFT point of view. We also emphasize
that  Eq.~(\ref{rentmp2}) is not obtained from (\ref{2}) by just
expressing $C$ and $C_2$ in terms of renormalized coupling
constants since these two low-energy constants are insufficient to absorb all
occuring divergences.
The renormalized couplings $C_R$ and $C_2^R$ depend on the
renormalization conditions through the renormalization group, but
this implicit dependence \emph{does not} cancel completely the
explicit dependence of the amplitude on $\mu$ and $\mu_n$.
However, different choices are equivalent up to the order of accuracy of the
calculation, provided that the chosen renormalization conditions respect power
counting.

We now turn to the KSW approach \cite{Kaplan:1998tg}, where the $T$-matrix
up to subleading order is given by\footnote{It results from Eq.~(\ref{2}) by
expanding in powers of $C_2$ and keeping the
first two terms.} 
:
\begin{equation}
T={C\over 1-C\,I^\Lambda(k)}+{2 k^2C_2+2CC_2I_3^\Lambda\over \left[
1-C\,I^\Lambda(k)\right]^2}. \label{chtmatrix}
\end{equation}
One can absorb \emph{all} divergences appearing in the above expression by
expressing the bare couplings $C$ and $C_2$ in terms of renormalized ones
$C_R (=C_R(\mu,\mu_3))$ and $C_2^R (=C_2^R(\mu,\mu_3))$.
The complete functional dependencies $C \equiv C(C_R, \, C_2^R )$ and $C_2 \equiv
C_2(C_R,
\, C_2^R )$ can be found in Ref.~\cite{Gegelia:1998iu}. For our purposes, it
is sufficient to expand these expressions in powers of $C_2^R$ which leads to
\begin{eqnarray}
C & = & {C_R\over 1+C_R\Delta(\mu)}+{2C_RC_2^R I_3^R(\mu_3) \over
\left[ 1+C_R\Delta(\mu)\right]^2}-{2C_RC_2^RI_3\over \left[
1+C_R\Delta(\mu)\right]^3}+\cdots , \label{npct3}\\
C_2 & = & {C_2^R\over \left[ 1+C_R\Delta(\mu)\right]^2}+\cdots .
\label{npct4}
\end{eqnarray}
In the KSW approach, the first term in Eq.~(\ref{npct3}) is treated
non-perturbatively while all
other terms in Eqs.~(\ref{npct3}) and (\ref{npct4}) are
taken into account perturbatively, order-by-order.
Substituting Eqs.~(\ref{npct3}) and (\ref{npct4}) into
(\ref{chtmatrix}) we obtain a finite renormalized expression:
\begin{equation}
\label{TKSW}
T={C_R\over 1-C_R\,I^R(\mu,k)}+{2C_RC_2^R\,I_3^R(\mu_3)\over
\left[ 1-C_R\,I^R(\mu,k)\right]^2}+{2 k^2C_2^R \over \left[
1-C_R\,I^R(\mu,k)\right]^2}+\cdots . \label{renchtmatrix}
\end{equation}
If we choose $\Delta_3\equiv I_3$ (i.e.~we set $\mu_3=0$),
the explicit dependence on $\mu$ can be completely compensated by implicit
dependence of running couplings $C_R$ and $C_{2}^R$ at any fixed order in the
EFT expansion. This is analogous to Refs.~\cite{Kaplan:1998tg,Kaplan:1998we} where
the dimensional regularization in combination with power
divergence subtraction (PDS) scheme is used.
Complete order-by-order renormalization scale-independence cannot be achieved
for any other choice of $\mu_3$.
Indeed, if the amplitude were $\mu$- and $\mu_3$-independent
order-by-order, the third term in
Eq.~(\ref{renchtmatrix}) should satisfy this condition by itself. Denoting
this term with $t_3$ we obtain
\begin{equation}
\sqrt{\frac{2\,k^2}{t_3}} =\left[ \frac{1}{C_R}-
I^R(\mu,k)\right]\,\frac{C_R}{\sqrt{C_2^R}}\,. \label{t3inv}
\end{equation}
As the integral $I^R(\mu,k)$ has an imaginary part which is renormalization
scale independent,
it follows from Eq.~(\ref{t3inv}) that $C_R/\sqrt{C_2^R}$ must be
renormalization scale independent. If this were the case, both $C_R$ and $C_2^R$
must be $\mu_3$-independent as it is easily seen from the real part of the
same equation.
However, in this case the explicit $\mu_3$-dependence of the second term in
Eq.~(\ref{renchtmatrix}) cannot be canceled by running of the
coupling constants {\rm $C_R$ and $C_2^R$}. We are forced to conclude that
the amplitude cannot be renormalization scale-independent
order-by-order.

It is instructive to  verify that the amplitude is indeed renormalization-scale
independent up to terms of order $\mathcal{O} ((C_2^R)^2)$.
By differentiating the expressions of the bare couplings in terms of renormalized ones, $C=C(C_R, \,
C_2^R)$ and $C_2=C(C_R, \, C_2^R)$ \cite{Gegelia:1998iu}, with respect to renormalization scales one obtains the
corresponding renormalization group equations for renormalized couplings.
For the beta-functions to first order in
$C_2^R$ these equations read:
\begin{eqnarray}
\frac{\partial C_R}{\partial\mu} & = & \frac{m}{2\,\pi^2}\,C_R^2
+\frac{m^2 \mu_3^3}{6\,\pi^4}\,C_R^2 C_2^R\,,\nonumber\\
\frac{\partial C_R}{\partial\mu_3} & = & \frac{m \mu_3^2}{\pi^2}\,C_R C_2^R\,,\nonumber\\
\frac{\partial C^R_2}{\partial\mu} & = & \frac{m}{\pi^2}\,C_R C_2^R\,,\nonumber\\
\frac{\partial C^R_2}{\partial\mu_3} & = & 0\,. \label{rengreqs}
\end{eqnarray}
We were unable to solve Eqs.~(\ref{rengreqs}) in a closed form but obtained
the expantion of the solution in powers of $C_2^R$:
\begin{eqnarray}
C_R(\mu,\mu_3) & = & \frac{2 \pi^2 C_R(0,0)}{2 \pi^2- \mu m C_R(0,0)
}+\frac{8 \pi^4 \mu _3^3 \, m\,  {C_R(0,0)} \,{C_2^R(0)}}{3
\left(2 \pi^2- \mu m C_R(0,0) \right)^3}+\cdots \,,\nonumber\\
C_2^R(\mu) & = & \frac{4 \pi^4 C_2^R(0)}{\left( 2 \pi^2- \mu m C_R(0,0)
  \right)^2}+\cdots\,. \label{solutions}
\end{eqnarray}
Substituting Eq.~(\ref{solutions}) into Eq.~(\ref{renchtmatrix}) leads to
\begin{eqnarray}
T & = & -\frac{2 \pi ^2 C_R(0,0)} { 2 \pi ^2 -C_R(0,0) \left[m
\mu +2 \pi^2 I^R(\mu,k)\right]}
 + \frac{8\,C_2^R(0)\,\pi ^4 k^2}{\left\{ 2 \pi ^2 -C_R(0,0)
\left[m \mu +2 \pi^2 I^R(\mu,k)\right]\right\}^2} \nn
&+& \cdots\,,
\label{rengrinvampl}
\end{eqnarray}
which is renormalization scheme independent up to the considered
order.

In case of the KSW approach, even using the PDS scheme, the residual
renormalization scale dependence is present in running coupling
constants if pions are included as explicit degrees of freedom.
The LO renormalized (running) coupling constant in the $^1S_0$ partial wave
given in Ref.~\cite{Kaplan:1998we} reads
\be
C_0^{(^1S_0)}(\mu)  =
-\frac{4\,\pi}{\mu\,m}\,\Biggl( \frac{1}{1- \left[\mu
\,\left(a+1/\Lambda_{NN}\right)
\right]^{-1}}+\frac{\mu}{\Lambda_{NN}}\Biggr)\,, \quad
\Lambda_{NN}  =  \frac{8 \pi \,f^2}{g_A^2
m}\,,\label{runningcoupling}
\ee
where $a$ is the $^1S_0$ scattering length,
$f$ denotes the pion decay constant normalized to be $f= 132 $ MeV
and $g_A$ is the axial-vector pion-nucleon coupling constant. In the expansion of
$C_0^{(^1S_0)}(\mu)$ in powers of $g_A$, the renormalization scale
dependence is present to all orders. This dependence is canceled
by an infinite number of higher-order terms in the expansion of the amplitude.
Notice that in KSW approach the appearance of the residual renormalization scale dependence
has a different origin compared to the Weinberg approach. There it arises from
the explicit renormalization scheme dependence of loop integrals due to
missing contributions of the corresponding higher-order renormalized coupling
constants.

To summarize the above considerations, renormalization scheme dependence is present
explicitly in loop contributions and implicitly in running
coupling constants. These two types of dependence exactly cancel each other
in the full amplitude. Due to the order-by-order calculations in
the KSW approach, there is a residual renormalization scheme
dependence generated by running couplings. On the other hand, in the case of
Weinberg's approach, the residual renormalization scale dependence arises from
loop contributions. There is no conceptual difference between the two
approaches from the point of view of the renormalization scale dependence.
If one approach is inconsistent, then so is the
other. In fact both are conceptually as consistent as perturbative
QCD is. Of course, the crucial issue is to choose an optimal
renormalization scheme (if such a scheme exists at all). In perturbative
QCD, when expressed in terms of the running coupling, the full
amplitudes are independent of the renormalization scale $\mu$.
Perturbative expressions are, however, $\mu$-independent only up to the
order of accuracy of the calculation. The residual $\mu$-dependence is canceled
by higher-order terms in perturbation theory. At high
energies one could choose renormalization scale much smaller than
the characteristic scale in a process under consideration. This would
generate a large value of the running coupling constant and, at the same time, lead
to large coefficients in the perturbative series. The failure of this kind
of perturbative scheme does not mean the failure of perturbation
theory in high-energy QCD in general. The only problem is that for
such an inappropriate choice of the renormalization scheme, ``higher-order''
$\mu$-dependent terms play a crucial role and are by no means
suppressed. EFT for few nucleons is conceptually
similar. Although observables are calculated by solving the corresponding
dynamical equations, one is still doing perturbative calculations with respect
to the chiral expansion. While the full amplitude is renormalization scheme
independent, truncated expressions at any finite order are generally not.

\medskip
We now switch to our next topic and consider Weinberg's approach based on a
finite cutoff rather than subtractive renormalization. In particular, we are
interested in the implications of taking the cutoff value very large.
To keep trace of the loop expansion, we first rewrite Eq.~(\ref{2}) by showing
explicitly factors of $\hbar$:
\begin{eqnarray}
T & = & \frac{C+C_2^2 \hbar I_5+ k^2C_2\left( 2-C_2 \hbar I_3\right)}{\left(
1-C_2 \hbar I_3\right)^2-\left[C+C_2^2 \hbar I_5+ k^2 C_2\left(
2-C_2 \hbar I_3\right)\right] \hbar I(k)}\,.
\label{2x}
\end{eqnarray}
The bare coupling constants $C$ and $C_2$ can be expressed in terms of the
scattering length $a$ and the effective range $r$ by matching the amplitude in
Eq.~(\ref{2x}) to the first two terms in the effective range expansion
 \be
\Re \left( T^{-1} \right) = - \frac{m}{4 \pi} \left( -\frac{1}{a} +
  \frac{1}{2} r k^2 + \ldots \right) \,,
\ee
which leads to the following expressions
\begin{eqnarray}
\label{bareLEC}
C & = & C(a, \, r , \, \Lambda ) =  \frac{6 \pi ^2 [ a^2 \hbar \Lambda^3  m (64 \hbar
-3 \pi  \Lambda r) -6 \left(D -3 \pi ^2 \Lambda
  m\right) -62 \pi  a \hbar\Lambda^2 m ]
}{5
\hbar \Lambda^2 m^2 \left[a^2 \hbar \Lambda^2 (16  \hbar -\pi \Lambda
 r)-12 \pi a \hbar \Lambda + 3 \pi ^2 \right]}\,, \nn
C_2 & = & C_2(a, \, r, \, \Lambda) = -\frac{6 \pi ^2
[ - D + a^2
 \hbar m \Lambda ^3 (16 \hbar - \pi r \Lambda )-12 \pi
   a \hbar m \Lambda ^2 + 3 \pi ^2 m \Lambda ]
}{\hbar  m^2 \Lambda ^4 \left[a^2 \hbar \Lambda ^2 (16
   \hbar - \pi r \Lambda )-12 \pi  a \hbar  \Lambda +3 \pi
   ^2\right]} \,,
\end{eqnarray}
with $D$ defined as
\be
D = \sqrt{3} \sqrt{\Lambda^2  m^2 (\pi -2
   a \hbar\Lambda)^2 \left(a^2 \hbar \Lambda^2 (16 \hbar -\pi
     \,\Lambda
  r)-12 \pi a \hbar \Lambda + 3 \pi ^2\right)}\,.
\ee
Substituting the resulting expressions for the bare couplings
$C(a, \, r, \, \Lambda)$ and $C_2(a, \, r, \, \Lambda)$ back into
Eq.~(\ref{2x}), we obtain  for the inverse
scattering amplitude
\begin{eqnarray}
T^{-1} &=& \frac{m }{4 \pi ^2 a
   \left[a \left(\pi  k^2 \,r\, \Lambda -4 \,\hbar\,
   \left(k^2+\Lambda ^2\right)\right)+2 \pi  \Lambda \right]}\,\Biggl\{2
\Lambda  \left[a^2 \,\hbar\, k^2 (\pi
   \,r\, \Lambda -4 \,\hbar)-2 \pi  a \,\hbar\,
   \Lambda +\pi ^2\right]\nonumber\\
&& +a \,\hbar\, k \ln
\frac{\Lambda -k}{\Lambda +k} \left[a \left(\pi
   k^2 \,r\, \Lambda -4 \,\hbar\, \left(k^2+\Lambda
   ^2\right)\right)+2 \pi  \Lambda \right]\Biggr\}+ i \hbar \frac{m k}{4 \pi }\,.
\label{invamplCutOff}
\end{eqnarray}
Although this expression for $T^{-1}$ possesses a finite limit as
$\Lambda \to \infty$ which, as desired, correctly reproduces the first two
terms in the effective range expansion (ERE),
\be
T^{-1} = - \frac{m}{4 \pi} \left( - \frac{1}{a} + \frac{1}{2} r k^2 - i \hbar\,
k \right) + \mathcal{O} \left(\Lambda^{-1} \right).
\label{invamplCutOffExpanded}
\ee
taking this limit without including all
relevant contributions of counterterms is a meaningless procedure within an EFT
\cite{Gegelia:gn}. Not surprisingly, one encounters pathologies, such as
e.~g.~the coupling $C_2$ becoming complex for positive values of the
effective range, see also Ref.~\cite{Beane:1997pk}.

To further explore the large-cutoff limit, we expand the
obtained expressions for the bare LECs in  Eq.~(\ref{bareLEC}) in powers of
$\hbar$ which leads to
\begin{eqnarray}
C & = &  \frac{4 \pi a}{m}+ \hbar \, \frac{3 a^4 r^2 \Lambda ^5 + 40 a^3 r
   \Lambda ^3+240 a^2 \Lambda }{30 m}+{\cal O}(\hbar^2)\,,\nonumber\\ [6pt]
C_2 & = & \frac{\pi a^2 r}{m}+ \hbar\, \frac{a^4
  r^2 \Lambda ^4 + 16 a^3 r \Lambda ^2 - 16
   a^2}{4 m \Lambda }+{\cal O}(\hbar^2)\,.
\label{CandC2expanded}
\end{eqnarray}
Making use of the standard splitting of bare quantities into renormalized ones
and counterterms,
\begin{eqnarray}
C & = & C^R +\sum_{k=1}^\infty \hbar^k\,\delta C_k\,,\nonumber\\
C_2 & = & C_2^R +\sum_{k=1}^\infty \hbar^k\,\delta C_{2 k}\,,
\end{eqnarray}
we identify the renormalized low-energy constants in this particular
scheme with \be \label{LECSrenormalized} C_R  =  \frac{4 \pi
a}{m}\,, \quad \quad C_2^R = \frac{\pi a^2 r}{m}\,.
\label{rencouplings} \ee Inverting the above expressions and
replacing in Eq.~(\ref{invamplCutOff}) the scattering length and the
effective range by $C_R$ and $C_2^R$, the inverse amplitude can be
re-written in terms of renormalized coupling constants. To see what
does the $\Lambda\to\infty$ limit correspond to in the language of
the EFT diagrams, we regard the loop expansion of the amplitude
(thus we reproduce the perturbative series summed up by iterating
the LS equation):
\begin{eqnarray}
T & = &C_R + 2C_2^R k^2
- i \hbar \frac{m\,k}{4 \pi} \left( C_R + 2 C_2^R k^2 \right)^2 \nn
&+& \hbar \frac{2 m\,k^4}{\pi^2} \Big[ - \left(C_2^R \right)^2 \Lambda +
\left(C_2^R  \right)^2 k^2 \Lambda^{-1} + C_2^R C_R \Lambda^{-1} + \mathcal{O}
\left(\Lambda^{-2} \right) \Big]  +\cdots 
\,,
\label{loopexp}
\end{eqnarray}
where ellipsis refer to higher-order terms in the loop expansion.
For momenta $k\gtrsim 1/a$ we cannot truncate the loop expansion in
Eq.~(\ref{loopexp}) at any finite order (i.e. in the language of
Feynman diagrams we need to sum up an infinite number of them). For
our demonstrating purposes we consider here the case when $r \ll a$.
With renormalized couplings of Eq.~(\ref{rencouplings}) the small
parameter of the EFT expansion is given by $k^2/m_s^2 \sim 2 C_2^R
k^2/C_R= a r k^2/2$, i.e. the hard scale of the problem is $m_s \sim
1/\sqrt{r a}$.\footnote{For very large scattering length $a \to
\infty$
a more suitable renormalization scheme should be used
\cite{Kaplan:1998tg,Gegelia:gn}.} The term linear in $\Lambda$ in
the second line of Eq.~(\ref{loopexp}) violates the dimensional
power counting and, for very large values of $\Lambda$ (much larger
than {\bf $m_s$}), yields the numerically dominant contribution at
one loop order, instead of being absorbed into redefinition of
higher-order coupling constants or, equivalently, being subtracted.
The situation is similar at higher orders in the loop expansion.
Hence, one completely looses the power counting which the EFT is
based on: terms which are supposed to be subtracted yield dominant
contributions to the amplitude. On the other hand, if $\Lambda$ is
taken of the order of the hard scale in the problem,  i.e.
$\Lambda \sim m_s$, the term linear in $\Lambda$ in the second line of
Eq.~(\ref{loopexp}) appears to be of order three and is beyond the
accuracy of the considered calculation. Notice that reducing the
value of $\Lambda$ considerably below the hard scale would lead to
large cutoff artefacts generated by terms with negative powers of
$\Lambda$. One is, therefore, forced to conclude that in the cutoff
theory, $\Lambda$ should ideally be chosen of the order of the hard
scale in the problem.

\section{Cutoff EFT: Renormalization versus ``peratization''}
\def\theequation{\arabic{section}.\arabic{equation}}
\setcounter{equation}{0}
\label{cutoffEFT}

In this section we further explore and extend the above ideas by considering
effective theory for an exactly solvable quantum mechanical model for
two nucleons interacting via the long- and short-range forces. This may be
regarded as a toy model for chiral EFT in the two-nucleon sector. We employ
both the subtractive renormalization and the cutoff formulation of the
resulting effective theory and discuss the similarities and differences
between these two approaches. We also explore the consequences of taking very
large values of the cutoff in this model.

\subsection{The model}
\label{ToyModel}
We consider two nucleons in the spin-singlet S-wave interacting
via the two-range separable potential
\be
\label{Vunderlying}
V (p ,\, p') =  v_{l} \, F_{l}(p)\,   F_{l}(p')+
  v_{s}\, F_{s}(p)\,   F_{s}(p')\,, \quad
F_l (p) \equiv \frac{\sqrt{p^2 + m_s^2}}{p^2 + m_l^2}\,,  \quad
F_s (p) \equiv \frac{1}{\sqrt{p^2 + m_s^2}}\,,
\ee
where the subscripts $l$ and $s$ refer to long- and short-range
interactions and the mass scales $m_l$ and $m_s$ fulfill the condition $m_l \ll
m_s$. Further, the dimensionless quantities
$v_l$ and $v_s$ denote the strengths of the long- and short-range
interactions, respectively. Our choice of the explicit form of $F_{l,s}(p)$
is entirely motivated by the simplicity of calculations. The reader may verify
that all conclusions reached in this section remain valid if one chooses,
for example, $F_{l,s}(p) \propto 1/(p^2 + m_{l,s}^2 )$. In this case, however,
one will need to go to subleading order in the EFT expansion in order to
explore the ``peratization'' procedure which would make the calculations
considerably more involved. 

For an interaction of a separable type, the off-shell
T-matrix can be easily calculated analytically by solving the corresponding
Lippmann-Schwinger  equation
\be
\label{LS}
T (p ,\, p'; \, k ) =  V (p ,\, p') + 4 \pi  \int \frac{l^2 dl}{(2 \pi)^3} V
(p ,\, l)
\frac{m}{k^2-l^2 + i \epsilon} T (l ,\, p'; \, k )\,,
\ee
where $m$ is the nucleon mass and $k$ corresponds to the on-shell momentum
which is related to the two-nucleon center-of-mass energy via $E_{\rm CMS} =
k^2/m$.
The phase shift $\delta (k)$ can be obtained from the on-the-energy-shell T-matrix
elements via
\be
T (k ,\, k; \, k ) = - \frac{4 \pi}{m} \frac{1}{k \cot  \delta (k)
  - i k}\,.
\ee
Here and in the following, we are particularly interested in the coefficients
entering the effective range expansion
\be
\label{ERE}
k  \cot  \delta (k)  = - \frac{1}{a} + \frac{1}{2}r k^2 + v_2 k^4 + v_3 k^6 +
\ldots\,,
\ee
with $a$, $r$ and $v_i$ referring to the scattering length, effective range and
the so-called shape parameters. The coefficients in the ERE generally scale
with the mass corresponding to the long-range interaction which gives rise to
the first left-hand cut in the T-matrix. Notice that the scattering
length can be tuned to any value by adjusting the strength of the
interaction. The coefficients in the ERE
can be expanded in powers of $m_l/m_s$ leading to the ``chiral'' expansion:
\bea
\label{EREexpanded}
a &=&  \frac{1}{m_l} \bigg( \alpha_a^{(0)}  + \alpha_a^{(1)} \frac{m_l}{m_s} +
\alpha_a^{(2)} \frac{m_l^2}{m_s^2} + \ldots  \bigg) \,, \nn
r &=& \frac{1}{m_l} \bigg( \alpha_r^{(0)}  + \alpha_r^{(1)} \frac{m_l}{m_s} +
\alpha_r^{(2)} \frac{m_l^2}{m_s^2} + \ldots  \bigg) \,, \nn
v_i &=& \frac{1}{m_l^{2 i -1}} \bigg( \alpha_{v_i}^{(0)} + \alpha_{v_i}^{(1)}
\frac{m_l}{m_s} +
\alpha_{v_i}^{(2)} \frac{m_l^2}{m_s^2} + \ldots  \bigg) \,,
\eea
where $\alpha_a^{(m)}$,  $\alpha_r^{(m)}$ and  $\alpha_{v_i}^{(m)}$ are
dimensionless constants whose values are determined by the specific form of the
interaction potential. We fine tune the strengths of the long- and short-range
interactions in such a way that they generate scattering lengths of a natural
size. More
precisely, we require that the scattering length takes the value $a =
\alpha_l/m_l$
($a = \alpha_s/m_s$) with a dimensionless constant $| \alpha_l | \sim 1$ ($|
\alpha_s  | \sim 1$)
when the short-range (long-range) interaction is switched off.  This leads to
\be
\label{strengths}
v_l = -\frac{8 \pi  m_l^3 \alpha _l}{m \left(\alpha _l m_s^2+m_l^2 \alpha _l-2
    m_s^2\right)}\,,
\quad
v_s = -\frac{4 \pi  m_s \alpha _s}{m \left(\alpha _s-1\right)}\,.
\ee
One then finds the following expressions for the first three terms in the
``chiral'' expansion in Eq.~(\ref{EREexpanded}).
\begin{itemize}
\item
Scattering length:
\be
\label{LET1}
\alpha_a^{(0)} =  \alpha_l \,, \quad  \quad
\alpha_a^{(1)} =  (\alpha_l -1 )^2 \alpha_s \,, \quad \quad
\alpha_a^{(2)} =  (\alpha_l -1 )^2 \alpha_l \alpha_s^2 \,.
\ee
\item
Effective range:
\bea
\label{LET2}
\alpha_r^{(0)} &=&  \frac{3 \alpha_l - 4}{\alpha_l} \,, \quad \quad
\alpha_r^{(1)} =  \frac{2 \left(\alpha _l-1\right) \left(3 \alpha
    _l-4\right) \alpha _s}{\alpha _l^2} \,,\nn
\alpha_r^{(2)} &=& \frac{\left(\alpha _l-1\right) \left(3 \alpha _l-4\right)
  \left(5 \alpha _l-3\right) \alpha _s^2+\left(2-\alpha _l\right) \alpha
   _l^2}{\alpha _l^3} \,.
\eea
\item
First shape parameter:
\bea
\label{LET3}
\alpha_{v_2}^{(0)} &=&  \frac{\alpha_l - 2}{2 \alpha_l} \,,\quad \quad
\alpha_{v_2}^{(1)} = \frac{\left[\alpha _l \left(13 \alpha
      _l-36\right)+24\right] \alpha _s}{4 \alpha _l^2} \,, \nn
\alpha_{v_2}^{(2)} &=&\frac{\left\{\alpha _l \left[\alpha _l \left(46 \alpha
        _l-159\right)+174\right]-60\right\} \alpha _s^2-4 \left(\alpha
    _l-2\right) \alpha _l^2}{4
   \alpha _l^3}\,.
\eea
\item
Second shape parameter:
\bea
\label{LET4}
\alpha_{v_3}^{(0)} &=& 0\,, \quad \quad
\alpha_{v_3}^{(1)} = \frac{\left(\alpha _l-2\right) \left(3 \alpha _l-4\right)
  \alpha _s}{2 \alpha _l^2} \,, \nn
\alpha_{v_3}^{(2)} &=& \frac{\left(3 \alpha _l-4\right) \left[\alpha _l \left(25
      \alpha _l-68\right)+40\right] \alpha _s^2-4 \left(\alpha _l-2\right)
  \alpha _l^2}{8 \alpha _l^3} \,.
\eea
\end{itemize}
Notice that the ERE is not applicable in the case $\alpha_l \to 0$ as follows
immediately from the considerations based on the Born approximation. We further
stress that in our model the leading terms in the $m_l/m_s$-expansion of the ERE
coefficients are completely fixed by the long-range interaction. The scenario
realized corresponds to a strong (at momenta $k \sim
m_l$) long-range interaction which needs to be treated non-perturbatively and
a weak short-range interaction which can be taken into account
perturbatively. We, however, emphasize that this particular hierarchy is not
important for our purposes.

\subsection{KSW-like approach and the low-energy theorems}
\label{LET}
Various coefficients in the ERE are \emph{correlated} with each other as a
consequence of the long-range interaction. In the context of effective (field)
theory, such correlations are to be regarded as low-energy theorems.
They have been discussed for the realistic case of nucleon-nucleon
interaction within the KSW scheme in
Refs.~\cite{Cohen:1998jr,Cohen:1999iaa,Cohen:1999ds} and were
shown to fail badly in the $^1S_0$ and $^3S_1$--$^3D_1$ channels. This failure
is a clear signal towards the non-perturbative nature of the one-pion exchange
in these channels. At the qualitative level, the low-energy theorems in the
pionful EFT simply reflect the hierarchy $M_\pi \ll \Lambda_{\rm hard}$
between the soft and hard scales in the problem, which set the upper bounds for
the convergence radii of the
ERE and chiral expansion, respectively. We will specify the precise meaning of the
low-energy theorems for the case at hand in the following.

\begin{figure}[tb]
  \begin{center}
\includegraphics[width=\textwidth,keepaspectratio,angle=0,clip]{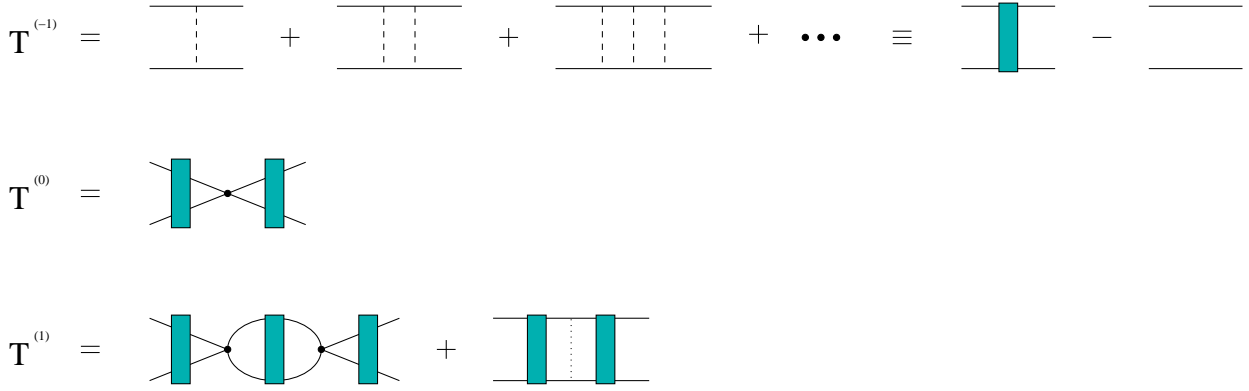}
\vskip 0.5 true cm
    \caption{Leading, next-to-leading and next-to-next-to-leading order
      contributions to the scattering amplitude in the KSW-like approach. The
      solid lines denote nucleons while the dashed ones represent an insertion
      of the lowest-order long-range interaction. Solid dots (dotted lines)
      denote an insertion of the lowest-order contact interaction $\propto
      C_0$ (subleading
      contribution to the long-range interaction).
    \label{fig1}
 }
  \end{center}
\end{figure}

We now develop EFT for the model specified above by keeping the long-range
interaction and replacing the short-range potential by a series of contact
zero-range interactions:
\be
V_{\rm short} (p, \, p' ) = C_0 + C_2 (p^2 + {p'} ^2) + \ldots \,,
\ee
where $C_{2n}$ are low-energy constants.
We begin with the most convenient and elegant
formulation which respects the standard dimensional power counting. To achieve
that we use subtractive renormalization for all divergent integrals and choose
the subtraction constant $\mu \sim m_l$. We expand the long-range
interaction in powers of $p/m_s$ in order to prevent the appearance of
positive powers of the large scale in the expressions for renormalized loop
diagrams which
would spoil the power counting. We also expand the strength of the long-range
interaction $v_l$ in Eq.~(\ref{strengths}) in powers of $m_l/m_s$ although
this is not necessary to maintain the power
counting. Here and in the following, we refer to
this approach as KSW-like. To be specific, we compute
the first few terms in the $Q/\lambda$-expansion of the T-matrix with $Q=\{k,
m_l , \mu \}$ and $\lambda = \{ m_s, m \}$. Notice that the natural size of
the short-range effects in our model suggests the scaling of the short-range
interactions in agreement with the naive dimensional analysis, i.e.~$C_{2n}
\sim Q^0$. The leading contribution to the T-matrix at order $Q^{-1}$ is
generated by the leading term in the $Q/\lambda$-expansion of the long-range
interaction
\bea
\label{long}
V_{\rm long} (p, \, p' ) &=&  v_{l} \, F_{l}(p)\,   F_{l}(p') \\
&\simeq&-
\frac{8 \pi  m_l^3 \alpha _l}{m \left(\alpha _l-2\right) ( p^2 + m_l^2)( {p
    '}^2 + m_l^2)}
\left[ 1 -\frac{\alpha _l m_l^2 }{\left(\alpha _l-2\right) m_s^2}
+ \frac{p^2}{2 m_s^2} + \frac{{p '}^2}{2 m_s^2} + \mathcal{O} \left(
  \frac{Q^4}{\lambda^4}
\right)\right] \nonumber
\eea
which scales as $Q^{-1}$ and, therefore,
needs to be summed up to an infinite order, see Fig.~\ref{fig1}.
This leads to the following expression for the on-the-energy shell T-matrix:
\be
T^{(-1)}=-\frac{8 \pi  m_l^3 \alpha _l}{m \left(k-i m_l\right){}^2 \left[k^2
    \left(\alpha _l-2\right)+2 i k m_l \left(\alpha _l-2\right)+2
    m_l^2\right]} \,,
\ee
from which one deduces
\bea
k \cot \delta &=& - \frac{4 \pi}{m} \frac{1}{T^{(-1)}} + i k \nn
&=& {} -\frac{m_l}{\alpha _l} + \frac{ \left(3
    \alpha _l-4\right)}{2 m_l \alpha _l} k^2  +
\frac{\left(\alpha _l-2\right)}{2 m_l^3 \alpha _l} k^4 \,.
\eea
Not surprisingly, one observes that the leading terms in the expansion of the ERE
coefficients in Eq.~(\ref{EREexpanded}) are correctly reproduced.
The first correction to the scattering amplitude at order $Q^0$ is given by
the leading-order contact interaction dressed with the iterated leading
long-range  interaction as visualized in
Fig.~\ref{fig1}. One finds
\be
T^{(0)} = \frac{C_0 \left(k+i m_l\right){}^2 \left[k^2 \left(\alpha
      _l-2\right)+2 m_l^2  \left(\alpha _l-1\right)\right]{}^2}{\left(k-i
    m_l\right){}^2 \left[k^2 \left(\alpha
   _l-2\right)+2 i k m_l \left(\alpha _l-2\right)+2 m_l^2\right]{}^2}.
\ee
Notice that all integrals entering $T^{(-1)}$ and $T^{(0)}$ are finite.
The effective range function $k \cot \delta$ at NLO
can be computed via
\be
k \cot \delta = - \frac{4 \pi}{m}  \frac{1}{T^{(-1)}} \bigg( 1 -
\frac{T^{(0)}}{T^{(-1)}}  \bigg) +  i k \,.
\ee
The ``chiral'' expansion of the coefficients in the ERE results from
expanding the right-hand side in this
equation in powers of $k^2$ and, subsequently, in powers of $m_l$. The
LEC $C_0$ can be determined from matching to $\alpha_a^{(1)}$ in
Eq.~(\ref{LET1}) which yields
\be
\label{C0_LO}
C_0 = \frac{4 \pi \alpha_s}{m m_s}\,.
\ee
This leads to the following predictions for $r$,  $v_2$ and $v_3$:
\bea
r^{\rm NLO}&=&\frac{1}{m_l} \Bigg[\frac{3 \alpha_l-4}{\alpha _l}
+\frac{2 \left( \alpha_l-1\right) \left(3 \alpha _l-4\right)
\alpha _s}{\alpha _l^2 m_s} m_l \Bigg]\,,\nn
v_2^{\rm NLO}&=&\frac{1}{m_l^3} \Bigg[ \frac{\left(\alpha _l-2\right)}{2
  \alpha _l}+
\frac{\left(\alpha _l \left(13 \alpha _l-36\right)+24\right) \alpha _s}
{4\alpha _l^2
   m_s} m_l \Bigg] \,, \nn
v_3^{\rm NLO}&=&\frac{1}{m_l^4} \, \frac{\left(\alpha _l-2\right) \left(3
    \alpha _l-4\right) \alpha _s}{2 \alpha _l^2 m_s} \,.
\eea
One observes that  $\alpha_r^{(1)}$, $\alpha_{v_2}^{(1)}$ and
$\alpha_{v_3}^{(1)}$ are correctly
reproduced at NLO. Using dimensional analysis it is easy to verify
that, in fact, $\alpha_{v_i}^{(1)}$ for
all $i$ \emph{must} be reproduced correctly at this order.

Finally, at next-to-next-to-leading order (NNLO) one has to
take into account the leading corrections to the long-range potential in
Eq.~(\ref{long}) and the contribution due to once
iterated leading-order contact term. Clearly, these contributions  have to be
dressed
by the iterated leading long-range interaction, see Fig.~\ref{fig1}.
The contribution $\propto C_0^2$ involves a linearly divergent integral which
we regularize with a cutoff $\Lambda \gg m_l$:
\be
\label{defenitionI1}
I_1^{\rm reg} \equiv 4 \pi m \int_0^\Lambda \frac{l^2 dl}{(2 \pi)^3}
\frac{1}{k^2 - l^2 + i \epsilon} = - \frac{m \Lambda}{2 \pi^2} - i \frac{m k
}{4 \pi}  + \mathcal{O} (\Lambda^{-1}) \,.
\ee
We carry out renormalization by subtracting the divergent
part of the integral $- m/(2 \pi^2) \int_\mu^\Lambda  dl$,  taking the limit $\Lambda
\to \infty$,
\be
I_1^{\rm reg} \to I_1^{\rm subtr} =  - \frac{m \mu}{2 \pi^2} - i \frac{m k
}{4 \pi}  \,,
\ee
and replacing the bare $C_0$ by the renormalized one $C_0 (\mu)$.  As already
pointed out before, we choose $\mu \sim m_l$ in order to be consistent with
the standard power counting based on the dimensional analysis. Clearly, the
above  procedure is exactly
equivalent to the power divergence subtraction prescription utilized in the
KSW framework. A simple calculation yields the following result for the
sub-subleading contribution to the amplitude:
\bea
T^{(1)} &=& \frac{\left(k+i m_l\right){}^2}{4 \pi ^2 m m_s^2 \left(k-i
    m_l\right){}^2 \left[k \alpha _l \left(k+2 i m_l\right)-2 \left(k+i
   m_l\right){}^2\right]{}^2} \Bigg[
-32 \pi ^3 k^2 m_l^3 \left(\alpha _l-2\right) \alpha _l \nn
&& {}+ \left( C_0 (\mu)\right) ^2 m^2 m_s^2 \left[k^2 \left(\alpha
    _l-2\right)+2 m_l^2
    \left(\alpha _l-1\right)\right]{}^2 \\
&& {} \times \frac{
\alpha _l \left[k^2 (-2 \mu
      -i
   \pi  k)+2 k (\pi  k-2 i \mu ) m_l+2 \pi  m_l^3\right]+2 (2 \mu +i
 \pi  k) \left(k+i m_l\right){}^2}{k \alpha _l \left(k+2 i
   m_l\right)-2 \left(k+i m_l\right){}^2} \Bigg].
\nonumber
\eea
The LEC $C_0( \mu )$ can be written in terms of the perturbative expansion as
follows
\be
C_0( \mu ) = C_0^{(0)} + C_0^{(1)}( \mu ) + \ldots \,,
\ee
where the superscript refers to the power of the soft scale $Q$. The first term
does not depend on $\mu$ and equals $C_0$ in Eq.~(\ref{C0_LO}).
The $\mu$-dependence of $C_0^{(1)} (\mu )$ can be determined by solving the
renormalization group equation
\be
\frac{d}{d \mu} \bigg[ T^{(-1)} + T^{(0)} + T^{(1)} \bigg] = 0\,.
\ee
One also needs one additional input parameter, such as e.~g.~$\alpha_a^{(2)}$,
in order to fix the
integration constant.  This leads to
\be
\label{C0_NLO}
C_0^{(1)} = \frac{8 \mu  \alpha _s^2}{m m_s^2}\,.
\ee
It is then easy to verify that the scattering amplitude $T^{(-1)} + T^{(0)} +
T^{(1)}$ is $\mu$-independent up to terms of order $Q^2$. Further, the
effective range function is given at this order by
\be
k \cot \delta = - \frac{4 \pi}{m}  \frac{1}{T^{(-1)}} \Bigg[ 1 -
\frac{T^{(0)}}{T^{(-1)}} +
\left( \frac{T^{(0)}}{T^{(-1)}} \right)^2   -
\frac{T^{(1)}}{T^{(-1)}}
 \Bigg] +  i k \,.
\ee
One then obtains the following predictions for the ERE coefficients:
\bea
r^{\rm NNLO}&=&\frac{1}{m_l} \Bigg[\frac{3\alpha _l -4}{\alpha _l}
+ \frac{2 \left(\alpha _l-1\right) \left(3 \alpha _l-4\right)
  \alpha _s}{\alpha _l^2 m_s} m_l \nn
&& {} +\frac{\left(\alpha _l-1\right) \left(3
      \alpha _l-4\right) \left(5 \alpha _l-3\right) \alpha _s^2+\left(2-\alpha
   _l\right) \alpha _l^2}{\alpha _l^3 m_s^2} m_l^2  \nn
&& {} -\frac{4 \mu m_l  \left(\alpha _l-1\right) \left(3 \alpha _l-4\right)
  \alpha _s^3 \left(\pi  m_l \left(3-5 \alpha _l\right)+4 \mu  \alpha
   _l\right)}{\pi ^2 \alpha _l^3 m_s^3}
+ \mathcal{O} \left( Q^4 \right)\Bigg] \,, \nn
v_2^{\rm NNLO}&=&\frac{1}{m_l^3} \Bigg[ \frac{\alpha _l-2}{2 \alpha _l}
+
\frac{\left(\alpha _l \left(13 \alpha _l-36\right)+24\right) \alpha _s}{4
  \alpha _l^2 m_s} m_l \nn
&& {}
+\frac{\left(\alpha _l \left(\alpha _l \left(46 \alpha
        _l-159\right)+174\right)-60\right) \alpha _s^2-4
   \left(\alpha _l-2\right) \alpha _l^2}{4 \alpha _l^3 m_s^2} m_l^2 \nn
&& {} + \frac{\mu  m_l \alpha _s^3 \left(\pi  m_l \left(\alpha _l \left(\alpha _l
        \left(46 \alpha _l-159\right)+174\right)-60\right)-2 \mu  \alpha _l
   \left(\alpha _l \left(13 \alpha _l-36\right)+24\right)\right)}{\pi ^2
 \alpha _l^3 m_s^3} \nn
&& {}
+ \mathcal{O} \left( Q^4 \right)\Bigg] \,,\nn
v_3^{\rm NNLO}&=&\frac{1}{m_l^5} \Bigg[
\frac{\left(\alpha _l-2\right) \left(3 \alpha _l-4\right) \alpha _s}{2
  \alpha _l^2 m_s} m_l + \frac{\left(3 \alpha _l-4\right) \left(\alpha _l
   \left(25 \alpha _l-68\right)+40\right) \alpha _s^2-4 \left(\alpha
   _l-2\right) \alpha _l^2}{8 \alpha _l^3 m_s^2} m_l^2 \nn
&& {} + \frac{\mu m_l  \left(3 \alpha _l-4\right) \alpha _s^3 \left(\pi  m_l
    \left(\alpha _l \left(25 \alpha _l-68\right)+40\right)-8 \mu  \left(\alpha
   _l-2\right) \alpha _l\right)}{2 \pi ^2 \alpha _l^3 m_s^3}
 + \mathcal{O} \left( Q^4 \right)\Bigg] \,,
\eea
where $Q = \{ m_l , \; \mu \}$.
As expected, the first three terms in the ``chiral'' expansion of \emph{all}
ERE coefficients  are reproduced correctly at NNLO. Notice further that the
contributions beyond the order of accuracy of the calculation are explicitly
renormalization-scale dependent, see section \ref{pionless} for a general
discussion.
The above results reveal
the meaning of the LETs in the present context. All $i$-th terms
$\alpha_x^{(i)}$ in the ``chiral'' expansion of the coefficients in the ERE, $x =
\{ a, \, r, \, v_2, \, \ldots \}$ are correlated with each other due to the
long-range interaction and its interplay with the short-range interaction in
the underlying model. The knowledge of $\alpha_{x_j}^{(i)}$ for one particular
$x_j$ is sufficient to predict $\alpha_{x_k}^{(i)}$ for all $k \neq j$. In an
EFT, short-range physics is incorporated in a
systematic way by taking into account contact interactions with an increasing
number of derivatives. Matching the strengths of the corresponding LECs to the
first $n$ terms in the ``chiral'' expansion of some of the ERE coefficients
allows to correctly describe the ``chiral'' expansion of
\emph{all} ERE coefficients up to order $m_l^n/m_s^n$. It should be emphasized
that at low energies and in the absence of external sources, the
appearance of the above mentioned correlations is \emph{the only} signature of the
long-range interaction in the 2N system.

\subsection{Weinberg-like approach with a finite cutoff}
\label{WeinbergCutoff}
\begin{figure}[tb]
  \begin{center}
\includegraphics[width=14cm,keepaspectratio,angle=0,clip]{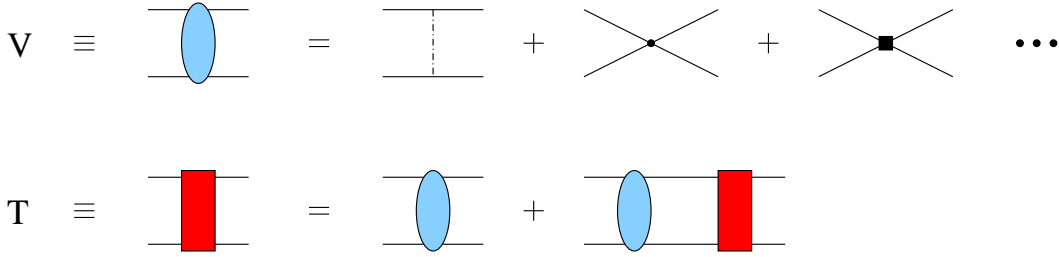}
\vskip 0.5 true cm
    \caption{Effective potential and scattering amplitude in the Weinberg-like
      approach. The dashed-dotted line refers to the full long-range
      interaction. Solid dot and filled rectangle refer to the leading and
      subleading contact interactions, respectively. For remaining notation
      see Fig.~\ref{fig1}.
    \label{fig2}
 }
  \end{center}
\end{figure}

An EFT formulation like the one described above which respects the
manifest power counting at every stage of the calculation is not
available in a general case of a long-range interaction which
is 
strong enough to have to be treated non-perturbatively such as e.~g.~the
one-pion exchange potential.
Here, one lacks a regularization prescription for \emph{all} divergent integrals
resulting from iterations of the potential in the LS equation
which would keep regularization artefacts small without, at the same time,
introducing a new hard
scale in the problem.
In the context of pionful EFT for few-nucleon systems,
the divergent
integrals are usually dealt with by introducing an UV cutoff $\Lambda$. In
order to keep regularization artefacts small, the cutoff, ideally, needs to be
taken of the order $\Lambda \sim m_s$ or higher. Clearly, this spoils the manifest
power counting for regularized loop contributions.\footnote{This, however,
  does not
  mean a breakdown of EFT since power counting is only required for
  \emph{renormalized} scattering amplitude.} We now consider the
Weinberg-like formulation in which the effective potential, given by the
long-range interaction and a series of contact terms, is iterated in
the LS equation to all orders, see the work by Lepage \cite{Lepage:1997} for a
related discussion. This is visualized in Fig.~\ref{fig2}.
We carry out renormalization by literally following
the steps outlined in Ref.~\cite{Lepage:1997} and summarized
in Ref.~\cite{Gasser:1982ap} in the following way: ''The
theory is fully specified by the values of the bare constants ...
once a suitable regularization procedure is chosen. In principle,
the renormalization program is straightforward: one calculates
quantities of physical interest in terms of the bare parameters at
given, large value of (ultraviolet cutoff) $\Lambda$. Once a
sufficient number of physical quantities have been determined as
functions of the bare parameters one inverts the result and
expresses the bare parameters in terms of physical quantities,
always working at some given, large value of $\Lambda$. Finally,
one uses these expressions to eliminate the bare parameters in all
other quantities of physical interest. Renormalizability
guarantees that this operation at the same time also eliminates
the cutoff.'' Notice that by iterating the truncated expansion for the
effective potential in the LS equation one unavoidably generates
higher-order contributions without being able to absorb all arising
divergences into redefinition of the LECs present in the considered truncated
potential. Thus, for the case at hand,
cutoff dependence in observables is expected to be eliminated only up to the
considered order
in the EFT expansion. We further emphasize that expressing the bare parameters
(i.~e.~LECs $C_i$) in terms of physical quantities is a non-trivial step due
to a nonlinear dependence of the scattering amplitude on $C_i$. The resulting
nonlinear equations may have no real solutions, especially when $\Lambda$ is
chosen to be considerably larger than $m_s$.
As pointed out in Ref.~\cite{Lepage:1997}, ``In fact, as nonlinearities
develop for small $a$'s,\footnote{In this work, $a$ refers to a
  coordinate-space cutoff, $a \sim \Lambda^{-1}$.} results often degrade, or,
in more extreme cases, the theory may become unstable or untunable''. The Wigner
bound in pionless EFT \cite{Phillips:1997xu} and the repulsive NN channels in the
pionful case \cite{PavonValderrama:2005wv,PavonValderrama:2005uj} may serve as examples for such an
untunable theory, see Ref.~\cite{Epelbaum:2006pt} for another example.
In the following, we consider the low-energy theorems for our model and
demonstrate explicitly that removing the
cutoff from the scattering amplitude by taking the limit $\Lambda \to \infty$
is not compatible with the EFT framework even if such a
limit exists  and the theory does not become untunable.

To be specific, we consider the effective potential of the form
\be
V_{\rm eff}^{(1)} (p ,\, p') =  v_{l} \, F_{l}(p)\,   F_{l}(p')
 + C_0 \,,
\ee
where the superscript of $V_{\rm eff}$ refers to the number of contact terms
included.
The off-shell T-matrix $T^{(1)} (p,\, p'; \, k)$ can be easily calculated by
solving the $2 \times 2$ matrix equation
\be
\label{LSmatrix}
t (k) =  v_{\rm eff} +  v_{\rm eff} \, \mathcal{G}(k) \,   t(k)
\ee
where we have defined
\be
V_{\rm eff}^{(1)} (p ,\, p') =\gamma^T (p) \, v_{\rm eff} \, \gamma (p '), \quad
T^{(1)} (p ,\, p', \, k) =\gamma^T (p) \, t (k) \, \gamma (p ')\,,
\ee
with
\be
v_{\rm eff} \equiv \left( \begin{array}{cc} v_l & 0 \\ 0 & C_0 \end{array}
\right)\,, \quad
\gamma ( p) \equiv \left( \begin{array}{c} F_l (p) \\ 1 \end{array}
\right)\,,  \quad
\mathcal{G}(k) \equiv \left( \begin{array}{cc} I_l(k) & I_{l1}^{\rm reg}(k) \\
    I_{l1}^{\rm reg} (k)
    &  I_1^{\rm reg} (k) \end{array}
\right)\,.
\ee
The integrals entering $\mathcal{G}(k)$ are given by
\bea
\label{otherintegrals}
I_l (k) &=& 4 \pi m \int_0^\infty \frac{l^2 \, dl}{(2 \pi)^3}
\frac{l^2+ m_s^2}{[k^2 - l^2 + i \epsilon][l^2 + m_l^2]^2} \nn
&=& \frac{m \left(-2 i k m_l+m_l^2+m_s^2\right)}{8 \pi  m_l \left(k+i
    m_l\right){}^2} \,, \nn
I_{l1}^{\rm reg} (k) &=& 4 \pi m \int_0^\Lambda \frac{l^2 \, dl}{(2 \pi)^3}
\frac{\sqrt{l^2+ m_s^2}}{[k^2 - l^2 + i \epsilon][l^2 + m_l^2]} \nn
&=& \frac{m}{2 \pi^2}
\Bigg( k \frac{\sqrt{k^2 + m_s^2}}{k^2 + m_l^2} \ln \Bigg(\frac{k + \sqrt{k^2
      + m_s^2}}{m_s} \Bigg) -
\frac{m_l \sqrt{m_s^2 - m_l^2}}{k^2 + m_l^2} \,
\text{arccot} \, \Bigg(
\frac{m_l}{\sqrt{m_s^2 - m_l^2}} \Bigg) \nn
&& {} +  \ln \left(\frac{m_s}{2 \Lambda }\right) - \frac{i \pi k
  \sqrt{k^2+m_s^2}}{2 \left(k^2+m_l^2\right)} \Bigg)  + \mathcal{O}
(\Lambda^{-1})\,,
\eea
and the integral $I_1^{\rm reg} (k)$ is defined in Eq.~(\ref{defenitionI1}).
Here and in what follows, we keep the cutoff $\Lambda$ at least of the
order $\Lambda \sim m_s$. We will, therefore, omit the finite
cutoff artefacts in order to keep the presentation
simple, i.~e.~we neglect the $\mathcal{O} (\Lambda^{-1})$-terms in
Eqs.~(\ref{defenitionI1}) and (\ref{otherintegrals}). The reader can easily
verify that taking into account finite cutoff  artefacts
(i.e. terms with negative powers of $\Lambda$) in the expressions for
regularized loop integrals does not alter the conclusions of this work.
With the above definitions, the LS equation (\ref{LSmatrix}) can be easily
solved leading to a somewhat lengthy expressions for the on-shell T-matrix
$T^{(1)}(k, \, k; \, k)$ which can be used to extract the coefficients in the
ERE.
One obtains for the scattering length
\be
\label{aWeinb1}
a^{(1)} = \frac{\pi  m_s \left\{C_0 m \left[2 \alpha _l \left(m_s \left(\Lambda
          -\text{s} m_l\right)+2 m_l^2 \ln (m_s/2 \Lambda )
\right)+\pi  m_l m_s\right]+4 \pi ^2 \alpha _l m_s\right\}}{m_l
\left\{2 \pi  m_s^2 \left(C_0 m \Lambda +2 \pi ^2\right)-C_0 m m_l
   \alpha _l \left[\text{s} m_s-2 m_l \ln (m_s/2 \Lambda)
       \right]^2\right\}} \,,
\ee
where we have introduced
\be
s \equiv 2 \frac{\sqrt{m_s^2-m_l^2}}{m_s}
\text{arccot}\left(\frac{m_l}{\sqrt{m_s^2-m_l^2}}\right)\,.
\ee
One can, in principle, determine the LEC $C_0$ by expanding $a^{(1)}$ in
powers of $m_l$
and matching the second term in this expansion to $\alpha_a^{(2)}$ as we did
in the case of the KSW-approach. However, in practice, ``chiral'' expansion of the
coefficients in the ERE is not available. We, therefore, determine $C_0$ for
a given value of the cutoff $\Lambda$, i.e.~$C_0(\Lambda)$, by
matching $a^{(1)}$ to the full expression of the scattering length
resulting in our model
\be
a_{\rm underlying} = \frac{m_l \left(2 \alpha _l-1\right) \alpha _s-\alpha _l
  m_s}{m_l \left(m_l \alpha _l \alpha _s-m_s\right)}\,,
\ee
which we regard as a synthetic data.
This leads to the following result for $C_0 (\Lambda)$:
\bea
\label{C0Weinberg}
C_0 (\Lambda) &=& 4 \pi ^3 \left(\alpha _l-1\right){}^2 m_s^2 \alpha _s \Bigg\{
m m_s \left[m_s \left(\pi -\text{s} \alpha _l\right)+2 m_l \alpha _l \ln
  \frac{m_s}{2 \Lambda }\right]{}^2 \nn
&&
-m \alpha _s \Bigg[m_s^2 \left(\alpha _l \left((\pi -\text{s}) m_l \left(-2
      \text{s} \alpha
   _l+\text{s}+\pi \right)
+2 \pi  \Lambda  \left(\alpha _l-2\right)\right)+2 \pi  \Lambda \right)
 \nn
&& {} +4
m_l^2 \alpha _l m_s \left((\pi -2
   \text{s}) \alpha _l+\text{s}\right) \ln \frac{m_s}{2 \Lambda
   }+4 m_l^3 \alpha _l \left(2 \alpha _l-1\right) \bigg( \ln
   \frac{m_s}{2 \Lambda } \bigg)^2\Bigg] \Bigg\}^{-1}\,.
\eea
Having determined the LEC $C_0 (\Lambda )$, we are now in the position to
verify the low-energy theorems by making predictions for the effective range
and shape coefficients. A straightforward calculation yields the following
result for \emph{renormalized} expressions for
$r^{(1)}$ and $v_2^{(1)}$:
\bea
\label{LETWeinberg}
r^{(1)} &=& \frac{1}{m_l} \Bigg[\frac{3 \alpha_l - 4}{\alpha _l}
+\frac{2 \left(\alpha _l-1\right) \left(3 \alpha _l-4\right) \alpha _s}{\alpha
  _l^2 m_s} m_l + \Bigg( \frac{4 \left(\alpha _l-2\right) \alpha _s }{\pi
  \alpha _l m_s^2} \left(\ln \frac{m_s}{2 \Lambda }+1\right) \nn
&& {} +
\frac{\left(\alpha _l-1\right) \left(3 \alpha _l-4\right) \left(5 \alpha
      _l-3\right) \alpha _s^2+\left(2-\alpha _l\right) \alpha
   _l^2}{\alpha _l^3 m_s^2} \Bigg) m_l^2 + \mathcal{O} \left( m_l^3
\right)\Bigg]\,,\nn
v_2^{(1)}&=&\frac{1}{m_l^3} \Bigg[ \frac{\alpha_l - 2}{2 \alpha _l}
+\frac{\left(\alpha _l \left(13
   \alpha _l-36\right)+24\right) \alpha _s}{4 \alpha _l^2
m_s} m_l
 + \Bigg(\frac{ \left(\alpha _l-2\right) \left(5 \alpha _l-6\right) \alpha
   _l^2 \alpha
  _s }{ \pi  \left(\alpha
   _l-1\right) \alpha _l^3 m_s^2} \left(\ln \frac{m_s}{2 \Lambda }+1\right)\nn
&& {} +
\frac{\left(\alpha _l \left(\alpha _l \left(46 \alpha
         _l-159\right)+174\right)-60\right) \alpha _s^2-4 \left(\alpha
   _l-2\right) \alpha _l^2}{4  \alpha _l^3 m_s^2} \Bigg) m_l^2
+ \mathcal{O} \left( m_l^3 \right)\Bigg]\,, \nn
v_3^{(1)}&=&\frac{1}{m_l^5} \Bigg[ \frac{\left(\alpha _l-2\right)
   \left(3 \alpha _l-4\right) \alpha _s}{2 \alpha _l^2 m_s} m_l
 + \Bigg(
\frac{2 \left(\alpha _l-2\right) \left(2 \alpha _l-3\right) \alpha _s
  }{\pi
   \left(\alpha _l-1\right) \alpha _l m_s^2}
\left(\ln \frac{m_s}{2 \Lambda } + 1 \right) \nn
&& {} +
\frac{\left(3 \alpha _l-4\right) \left(\alpha _l \left(25 \alpha
      _l-68\right)+40\right) \alpha _s^2-4 \left(\alpha _l-2\right) \alpha
  _l^2}{8
   \alpha _l^3 m_s^2}
 \Bigg) m_l^2  + \mathcal{O} \left( m_l^3 \right)\Bigg]\,.
 \eea
Again, not surprisingly, one observes that the subleading terms in the
``chiral'' expansion of the ERE coefficients are correctly reproduced. In fact,
any quantum-mechanically well-defined short-range interaction accompanied with
the underlying long-range force would do equally good job in describing
correlations between $\alpha_{x}^{(1)}$. While all coefficients
$\alpha_{x}^{(1)}$ have to be reproduced correctly at this order once the
short-range parameter entering the effective potential is appropriately tuned
(as guaranteed by the analytic structure of the scattering amplitude), there
is no restriction regarding higher-order terms in the ``chiral''
expansion.\footnote{A careful reader may realize that also the sub-subleading
  coefficients in the ``chiral'' expansion of $r$, $v_2$ and $v_3$ are
  correctly reproduced once the cutoff is tuned to the value $\Lambda = e
  m_s/2$. This, in fact, also holds true for higher coefficients in the ERE
  and can be traced back to the fact that in the case at hand, the cutoff
  $\Lambda$ itself may
  be considered as an additional short-range "counterterm'' provided one
  allows for a fine tuning of $\Lambda$. The resulting expressions are completely
  equivalent to the next-higher-order calculation in the Weinberg-like
  approach and provide explicit evidence for the validity of low-energy
  theorems in that case.}
Indeed, one observes that the coefficients $\alpha_{r, \, v_i}^{(2)}$ in
Eq.~(\ref{LETWeinberg}) deviate from their correct values given in
Eqs.~(\ref{LET2})-(\ref{LET4}).
Moreover, since the included LEC is insufficient to absorb all divergencies
arising from iterations of the LS equation, nothing prevents the appearance of
positive powers or logarithms of the cutoff $\Lambda$ in the expressions for
$\alpha_{r, v_i}^{(n)}$ with $n \geq 2$.\footnote{The appearance of
  only logarithmic dependence on $\Lambda$ in Eq.~(\ref{LETWeinberg}) is
  specific to the form of the long-range interaction and the order in the EFT
  expansion. We have verified that positive powers of $\Lambda$ occur in
  the expressions for $\alpha_{x}^{(3)}$ when one includes the
  subleading contact interaction in the effective potential.} The results
in Eq.~(\ref{LETWeinberg}) show that this is indeed the case. The
dependence on $\Lambda$ occurs, however, only in contributions beyond the
accuracy of calculation and, obviously, does not affect the predictive power
of the EFT provided the cutoff is chosen to be of the order of the
characteristic hard scale in the problem, $\Lambda \sim m_s$. Taking values
$\Lambda \gg m_s$ artificially enhances certain higher-order contributions in the
``chiral'' expansion of the ERE coefficients spoiling the predictive power of
the theory.

The appearance of positive powers of $\Lambda$ and/or logarithmic terms in the
predicted ``chiral'' expansion of the effective range
and the shape parameters in Eq.~(\ref{LETWeinberg}) may give the wrong
impression that no finite limit exists for $r^{(1)} (\Lambda )$ and $v_i^{(1)}
( \Lambda )$ as $\Lambda \to \infty$.  In fact, taking
the limit $\Lambda \to \infty$ does not commute with the Taylor expansion of
the ERE coefficients in
powers of $m_l$. It is easy to see, that all coefficients in the ERE as well
as the on-shell T-matrix approach a finite limit as $\Lambda \to \infty$.
Substituting the value for $C_0 (\Lambda )$ from Eq.~(\ref{C0Weinberg}) into
the solution of the LS equation (\ref{LSmatrix}) and taking the limit $\Lambda
\to \infty$ one obtains the following cutoff-independent result for the
inverse amplitude:
\bea
\label{Tperatized}
(T^{(1)}_{\rm peratized})^{-1} &=& i \frac{k m}{4 \pi}  - \frac{m}{8 \pi  m_l^3
  \left(k^2+m_s^2\right)
 \left(\alpha _l m_s+m_l \left(1-2 \alpha _l\right) \alpha _s\right)} \Big(
2 m_l^4 m_s^2 \left(m_s-m_l \alpha _l \alpha _s\right) \nn
&& {} + k^2 m_l^2 \left(\left(4-3 \alpha _l\right) m_s^3+m_l^2 \alpha _l
  m_s+m_l \alpha _s \left(\left(2 \alpha _l-3\right) m_s^2+m_l^2 \left(1-2
      \alpha
   _l\right)\right)\right) \nn
&& {} + k^4 \left(-\alpha _l m_s \left(m_l^2+m_s^2\right)-m_l \alpha _s
  \left(m_l^2 \left(1-2 \alpha _l\right)+m_s^2\right)+2 m_s^3\right)  \Big)\,.
\eea
The above procedure is very much in spirit of the so-called
peratization, the technique introduced by Feinberg and Pais
\cite{Feinberg:1963zz,Feinberg:1964zz}, see also \cite{Guttinger:1965zz}, 
to evaluate higher-order corrections to S-matrix in
non-renormalizable field theories. The essential idea of this method consists
in the resummation of the most divergent contributions to the Born series. 
In the late sixties of the last century, this technique was widely used in
potential scattering as an attempt to generate approximations to the
scattering length for different classes of singular potentials. In some cases
where the exact solution to the Schr\"odinger equation with a given singular
potential is known, peratization was indeed shown to provide reasonable
approximations to the scattering length while in other cases this approach
fails completely, see  \cite{Frank:1971xx} for a comprehensive review
article. 

The
cutoff-removed results for the ERE coefficients can be read off from
Eq.~(\ref{Tperatized}):
\bea
r_{\rm peratized}^{(1)} &=& \frac{m_l^3 \alpha _s+m_l^2 \left(\alpha
    _l-2\right) m_s+m_l \left(2 \alpha _l-3\right) m_s^2 \alpha _s+\left(4-3
    \alpha _l\right) m_s^3}{m_l
   m_s^2 \left(m_l \left(2 \alpha _l-1\right) \alpha _s-\alpha _l m_s\right)}
 \nn
&=& \frac{1}{m_l} \Bigg[\frac{3 \alpha_l - 4}{\alpha _l}
+\frac{4 \left(\alpha _l-1\right){}^2 \alpha _s}{\alpha _l^2 m_s} m_l
\nn
&& {} +
\frac{\alpha _l^3 \left(8 \alpha _s^2-1\right)+\alpha _l^2
   \left(2-20 \alpha _s^2\right)+16 \alpha _l \alpha _s^2-4 \alpha
   _s^2}{\alpha _l^3 m_s^2} m_l^2
+ \mathcal{O} \left( m_l^3 \right)\Bigg]\,,\nn
(v_2^{(1)})_{\rm peratized} &=&-\frac{\left(m_l^2-m_s^2\right){}^2
  \left(\left(\alpha _l-2\right) m_s+m_l \alpha _s\right)}{2 m_l^3 m_s^4
  \left(m_l \left(2 \alpha
   _l-1\right) \alpha _s-\alpha _l m_s\right)} \nn
&=& \frac{1}{m_l^3} \Bigg[ \frac{\alpha_l - 2}{2 \alpha _l} +
\frac{\left(\alpha _l-1\right){}^2 \alpha _s}{\alpha _l^2 m_s} m_l
 +\frac{\alpha _l^3 \left(2 \alpha _s^2-1\right)+\alpha _l^2 \left(2-5
   \alpha _s^2\right)+4 \alpha _l \alpha _s^2-\alpha _s^2}{\alpha _l^3 m_s^2}
m_l^2
\nn
&& {}
+ \mathcal{O} \left( m_l^3 \right)\Bigg]\,, \nn
(v_3^{(1)})_{\rm peratized} &=& \frac{\left(m_l^2-m_s^2\right){}^2
  \left(\left(\alpha _l-2\right) m_s+m_l \alpha _s\right)}{2 m_l^3 m_s^6
  \left(m_l \left(2 \alpha _l-1\right)
   \alpha _s-\alpha _l m_s\right)}  \nn
&=& \frac{1}{m_l^5} \Bigg[ -\frac{\alpha _l-2}{2 \alpha _l m_s^2} m_l^2 +
\mathcal{O} \left( m_l^3 \right)\Bigg]\,.
\eea
One observes that the results after removing the cutoff fail to reproduce the
low-energy theorems by yielding wrong values for $\alpha_{r}^{(1)}$ and
$\alpha_{v_i}^{(1)}$ (notice that, per construction, the scattering length
corresponding to $T^{(1)}_{\rm  peratized}$  exactly matches $a_{\rm underlying}$).
Note that we could allow for a stronger fine-tuning in our model to
make both the long- and short-range interactions nonperturbative at $k \sim
m_l$ (as it probably happens in the realistic case of NN scattering). The
breakdown of LETs in the ``peratized'' expressions would then imply that the
coefficients in the ERE are completely uncorrelated with each other, that is,
the predictive power of such an approach is the same as in the theory with
only short-range interactions (i.~e.~''pionless'' theory).

The breakdown of LETs in the ``peratized'' approach can be traced back to
spurious $\Lambda$-dependent contributions in the T-matrix which are
irrelevant (at the order of calculations) in the regime $\Lambda
\sim m_s$ but become numerically dominant if $\Lambda \gg m_s$. In general,
such spurious terms involve positive powers of $\Lambda$ which, as $\Lambda$
gets increased beyond the hard scale $m_s$, become,
at some point, comparable in size with the lower-order terms.
For example, as already mentioned before, terms linear in $\Lambda$ will show up in
the renormalized expressions for $\alpha_x^{(3)}$ at next-higher order. Low-energy
theorems will then break down as the cutoff will approach the scale $\Lambda
\sim m_s^2/m_l$.  The unavoidable appearance of ever higher power-law
divergences when going to higher orders in the EFT expansion implies that the
cutoff should not be increased beyond the hard scale in the problem, which
leads to the optimal
choice $\Lambda \sim m_s$.

\section{Summary and conclusions}
\label{summary}

We discussed some conceptual aspects of renormalization in the
context of effective field theories for the two-nucleon system.  First,
we considered renormalization scheme dependence of the scattering
amplitude in the KSW and Weinberg's approaches. Renormalization scale dependence
is present explicitly in the loop contributions and implicitly due to the running of the
coupling constants. These two types of dependence cancel exactly
in the full amplitude. Contrary to widespread belief, we showed that
renormalization scheme independence of the amplitude in pionless EFT based on
the KSW framework is only achievable up to the order to which the calculations are
performed. The residual renormalization-scheme dependence arises from the
running coupling constants. On the other hand, in the Weinberg's approach the
residual renormalization scale dependence is generated by loop
contributions. From this point of view, the KSW framework does not offer any
conceptual advantage over the Weinberg's approach. If one approach is
conceptually inconsistent, then so is the other. In fact both are conceptually
as consistent as perturbative QCD. Clearly, the crucial point is to choose the
appropriate renormalization condition.

Secondly, we considered the cutoff version
of pionless theory for NN scattering in the $^1S_0$ partial wave
up to next-to-leading order. We expressed the scattering amplitude in terms
of renormalized coupling constants and explored the consequences of
taking the cutoff $\Lambda$ very large, i.e.~much larger than the hard scale
in the problem. Making use of the loop expansion for the scattering amplitude,
we observed that the contributions which diverge in the
limit $\Lambda \to \infty$, instead of being
absorbed into redefinition of higher-order coupling constants (or,
equivalently, being subtracted), start playing a dominant role as $\Lambda$ is
increased significantly beyond the pertinent hard scale.
One, therefore, completely looses the power
counting (at the level of the amplitude) on which the EFT is based.
On the other hand, if $\Lambda$ is chosen of the order of the hard scale,
violation of the power counting by terms proportional to $\Lambda$
appears to be beyond the accuracy of the calculation.

To further explore the role of the cutoff we constructed
a toy-model for pionful EFT. Specifically, we developed an effective theory
for an exactly
solvable quantum mechanical problem with long- and short-range interactions of
a separable type. We revealed the meaning of low-energy theorems in this model
using the KSW-like framework with subtractive renormalization and demonstrated
their validity in the Weinberg-like approach with a finite cutoff $\Lambda$ as
long as it is chosen of the order of short-range scale. Taking the limit
$\Lambda \to \infty$ while keeping the scattering length at its correct value
yields a finite result for the amplitude but violates the
low-energy theorems. This procedure is, therefore, not compatible with the EFT
framework. It is much more in spirit of peratization
\cite{Feinberg:1963zz,Feinberg:1964zz,Guttinger:1965zz,Frank:1971xx} than
renormalization as it is understood in the context of EFT.
Contrary to popular opinion, the considered example demonstrates
that the existence of a finite limit of the amplitude as $\Lambda \to \infty$
under requirement, that certain low-energy observables such as e.~g.~the
scattering length are kept at their physical values, is not yet sufficient
for a proper renormalization in the context of
chiral EFT (neither is it necessary, see e.g.~\cite{Lepage:1997,Lepage:2000}).
We argue that $\Lambda$ should not be
increased (considerably) beyond the short-range scale in the problem in EFT
calculations of that kind.

\acknowledgments We would like to thank Dalibor Djukanovic,
Ulf-G.~Mei{\ss}ner, Daniel Phillips and Manuel Pav\'on Valderrama
for useful comments on the manuscript. The work of E.E.~was
supported by funds provided by the Helmholtz Association to the
young investigator group ``Few-Nucleon Systems in Chiral Effective
Field Theory'' (grant VH-NG-222) and to the virtual institute ``Spin
and strong QCD'' (VH-VI-231), by the DFG (SFB/TR 16 ``Subnuclear
Structure of Matter'') and by the EU HadronPhysics2 project ``Study
of strongly interacting matter''. J.G.~acknowledges the support of
the Deutsche Forschungsgemeinschaft (SFB 443) and Georgian National
Foundation grant GNSF/ST08/4-400.


\end{document}